\documentclass{article}
\usepackage{graphicx}%
\usepackage{multirow}%
\usepackage{amsmath,amssymb,amsfonts}%
\usepackage[top=1in, bottom=1.25in, left=0.9in, right=0.9in]{geometry}
\usepackage{amsthm}%
\usepackage{mathrsfs}%
\usepackage[title]{appendix}%
\usepackage{xcolor}%
\usepackage{textcomp}%
\usepackage{manyfoot}%
\usepackage{booktabs}%
\usepackage{acronym}
\usepackage{algorithm}%
\usepackage[numbers,sort&compress]{natbib}
\usepackage{algorithmicx}%
\usepackage{algpseudocode}%
\usepackage{siunitx}
\usepackage{listings}%
\usepackage{subcaption}
\usepackage{verbatim}
\usepackage{authblk}
\usepackage{comment}

\acrodef{qg}[QG]{Quantum Gravity}
\acrodef{dof}[DOFs]{degrees of freedom}
\acrodef{pmns}[PMNS]{Pontecorvo--Maki--Nakagawa--Sakata}
\acrodef{dm}[DM]{dark matter}
\acrodef{sm}[SM]{Standard Model}
\acrodef{gr}[GR]{General Relativity}
\acrodef{mc}[MC]{Monte Carlo}
\acrodef{cl}[C.L.]{confidence level}
\acrodef{wp}[WP]{wave packet}
\acrodef{qd}[QD]{quantum decoherence}
\title{\bf Potential of Neutrino Telescopes to Detect Quantum Gravity-Induced Decoherence in the Presence of Dark Fermions}
\author[1]{Alba Domi\footnote{\texttt{alba.domi@fau.de}}}
\author[1]{Thomas Eberl\footnote{\texttt{thomas.eberl@fau.de}}}
\author[2]{Dominik Hellmann\footnote{\texttt{dominik.hellmann@tu-dortmund.de}}}
\author[2]{Sara Krieg\footnote{\texttt{sara.krieg@tu-dortmund.de}}}
\author[2]{Heinrich P\"as\footnote{\texttt{heinrich.paes@tu-dortmund.de}}}
\affil[1]{Erlangen Centre for Astroparticle Physics, Friedrich-Alexander-Universit\"at Erlangen-N\"urnberg, Nikolaus-Fiebiger-Str. 2, 91058 Erlangen, Germany}
\affil[2]{Fakult\"at f\"ur Physik, Technische Universit\"at Dortmund, Otto-Hahn-Straße 4, 44227 Dortmund, Germany}
\begin{document}

\maketitle

\begin{abstract}
We assess the potential of neutrino telescopes to discover quantum-gravity-induced decoherence effects modeled in the open-quantum system framework and with arbitrary numbers of active and dark fermion generations, such as particle dark matter or sterile neutrinos. The expected damping of neutrino flavor oscillation probabilities as a function of energy and propagation length thus encodes information about quantum gravity effects and the fermion generation multiplicity in the dark sector. We employ a public Monte-Carlo dataset provided by the IceCube Collaboration to model the detector response and estimate the sensitivity of IceCube to oscillation effects in atmospheric neutrinos induced by the presented model.
Our findings confirm the potential of very-large-volume neutrino telescopes to test this class of models and indicate higher sensitivities for increasing numbers of dark fermions.
\end{abstract}

\begin{comment}
\begin{abstract}
    \ac{qg}-induced decoherence could provide a useful tool in the search for dark fermions. This effect is expected to influence neutrino oscillations, with its impact depending on both the number of active generations and the presence of additional dark fermions, such as \ac{dm} or sterile neutrinos, sharing the same unbroken gauge quantum numbers as neutrinos. In this work, we assess the potential of neutrino telescopes to detect QG effects in the presence of an arbitrary number of fermions. Our focus is on simulated atmospheric neutrinos observed by the IceCube Neutrino Observatory, using its publicly available \ac{mc} samples. Our findings show that the sensitivity to QG effects increases with the number of dark fermions considered.
\end{abstract}
\end{comment}

\section{Introduction}\label{sec:intro}

The most fundamental theories in physics, \ac{gr} and the \ac{sm} of particle physics, have been tested with very high precision.
However, unifying these theories into a universal theory of \acf{qg} remains one of the greatest challenges in fundamental physics~\cite{Addazi:2021xuf}.
Approaches like string theory or loop quantum gravity seek to integrate these theories into a more comprehensive framework,
in which \ac{gr} and Quantum Theory emerge as limiting cases, c.f. for example
Ref.s~\cite{Polchinski:1998rq, Becker:2006dvp, Ibanez:2012zz}
and~\cite{Rovelli:1987df,Rovelli:1989za,Rovelli:1994ge,Ashtekar:2004vs,Thiemann:2006cf,Celada:2016jdt}.
However, to date, no experimental evidence exists favoring any of these approaches to \ac{qg},
underscoring the profound difficulties all these approaches encounter~\cite{Kiefer:2013jqa}.
% Our approach diverges from these theories, focusing not on the underlying theory itself but rather on the observable effects of \ac{qg}.
Another possible avenue to study and find evidence for \ac{qg} is to pursue the bottom-up approach by
formulating a low energy effective theory of how \ac{qg} could impact another, known quantum system.
One way to formulate such an effective theory is by utilizing the open quantum system framework in which
we consider the equations of motion of a known quantum system and treat \ac{qg} as an environment.
As extensively explored in the literature~\cite{Hawking:1982dj,Ellis:1983jz,Ellis:1992dz,Huet:1994kr,Liu:1997zd,Chang:1998ea,Benatti:2000ph,Gago:2002na,Klapdor-Kleingrothaus:2000kdx,Hooper:2004xr,Hooper:2005jp,Anchordoqui:2005gj,Barenboim:2006xt,Stuttard:2020qfv},
\ac{qg} effects could induce decoherence, i.e. the loss of coherence in a quantum system due to its interaction and entanglement with an environment,
that is a or even the defining characteristic of the quantum-to-classical transition~\cite{Zeh:1970zz},
potentially leading to observable phenomena.
It follows that the exploration of quantum gravitational effects and the quantum-to-classical transition represents an exciting research frontier.

Applying this framework to the study of neutrino flavor oscillations, \ac{qg} induced decoherence could provide a useful tool in the search for unknown electrically neutral particles~\cite{Hellmann:2021jyz, Hellmann:2022cgt}, i.e. possible \ac{dm} candidates. 
This conjecture is based on the central assumption that \ac{qg} interactions maximally violate global quantum numbers—such as lepton flavor—while conserving only unbroken gauge quantum numbers, as discussed for example in Ref.s~\cite{Anchordoqui:2006xv,Witten:2017hdv,Harlow:2018jwu}.
This assumption is rooted in the so called No-Hair theorem~\cite{Hawking:1975vcx,Page:1980qm}:
Black holes only conserve energy, angular momentum and gauge charges, all other information of particles falling inside is lost.
If quantum black holes in the spacetime foam share this property, this implies that flavor information of a quantum system reduces while it interacts with this environment.
Furthermore, assuming the existence of new fermionic particle species being singlets with respect to the unbroken \ac{sm} gauge group $SU(3)_c \times U(1)_{\mathrm{em}}$,
this could induce transitions between neutrinos and these unknown fermion species that are otherwise forbidden or highly suppressed.
Candidates for such dark fermions include WIMPS~\cite{Roszkowski:2017nbc}, sterile neutrinos~\cite{Dasgupta:2021ies}, and FIMPS~\cite{Westhoff:2023xho}. 
Given that numerous searches for \ac{dm} candidates have not yet produced significant evidence, it is possible that if such new particles exist, they are entirely decoupled from the Standard Model in the current phase of the universe and hence interact purely gravitationally with known matter.
Therefore, besides providing a potential window on \ac{qg}, the proposed mechanism offers a rare and valuable opportunity to search for these highly elusive particles.

A further consequence of the maximally flavor violating properties of \ac{qg} is that information distinguishing particles of different
flavor would degrade over time, resulting in fully democratic transition probabilities~\cite{Hellmann:2021jyz} across all
generations of active neutrinos as well as the hypothetical dark fermions.
Hence, an initially pure neutrino system would always develop a dark component after travelling a sufficiently long distance.
This, in turn, results in distinct oscillation signatures in neutrino oscillation experiments, reflecting the influence of the dark fermions.
Consequently, long baseline neutrino experiments could be sensitive to such an effect, allowing to probe the \ac{qg} parameters~\cite{Gomes:2020muc,Coloma:2018idr, Lessing:2023P2,QD_review_exp,ICECUBE:2024fej,Domi:2024ypm}.
Neutrino telescopes such as IceCube~\cite{IC_overview}, ANTARES~\cite{antares} and KM3NeT~\cite{KM3Net:2016zxf} have a significant advantage over the other neutrino experiments, as they can exploit a wider range of $L\,/\,E$ ratios, allowing for high sensitivity to a broader region of the \ac{qg} parameter space.
Currently, no evidence for \ac{qg}-induced quantum decoherence has been observed~\cite{Coloma:2018idr, Lessing:2023P2,QD_review_exp,ICECUBE:2024fej}, and the most stringent limits on decoherence parameters growing with increasing energy have been established by the IceCube collaboration using high energy atmospheric neutrinos \cite{ICECUBE:2024fej}. 

Meanwhile, the KM3NeT infrastructure is under construction in the Mediterranean Sea and features two detectors, ORCA and ARCA, each designed for studying neutrinos at different energy ranges. ORCA is optimized for detecting neutrinos with energies starting from \SI{1}{\giga\electronvolt}, where matter effects in the Earth enhance its sensitivity to high-precision measurements of neutrino oscillations using atmospheric neutrinos.
This feature allows ORCA to probe decoherence parameters decreasing with growing neutrino energy with high sensitivity.
On the other hand, ARCA is designed for high-energy neutrino studies, extending to PeV energies and beyond, and offers complementary sky coverage to IceCube. Together, these experiments provide a comprehensive probe of neutrino behavior across a wide range of energy scales.

In this work, the IceCube Neutrino Observatory is of particular interest due to its extensive public dataset release, which allows to investigate the detector's potential for testing the proposed model.
We perform a statistical analysis on a pseudo data set generated according to the standard three neutrino oscillation hypothesis to investigate the potential constraints that IceCube could impose on \ac{qg} parameters using atmospheric neutrinos, considering different total numbers of dark fermions and various decoherence models. Our work extends the previous IceCube analysis on \ac{qd} \cite{ICECUBE:2024fej} by including a varying number of dark fermions, allowing for a generalization; when the number of dark fermions is set to zero, our results align with those of IceCube.
Furthermore, the assumption that \ac{qg} is maximally flavor violating corresponds to the \textit{state selection} models discussed in Ref.~\cite{ICECUBE:2024fej} as these models also lead to the uniform flavor distribution in the \ac{qd} limit.
To provide a realistic estimate of the potential sensitivity of IceCube to the model parameter space, we also take into account systematic effects described in reference~\cite{ICECUBE:2024fej} and references therein.
%For a more realistic and conservative result, \com{all} systematic uncertainties need to be included, which could motivate a future full analysis involving both the IceCube and KM3NeT detectors.

In section~\ref{sec:deco_model}, we begin by explaining the phenomenological description of \ac{qg} induced decoherence for a neutrino system, followed by an analysis of the potential to constrain the decoherence parameters for the different models in section~\ref{sec:analysis}.
Finally, we summarize and discuss our findings in section~\ref{sec:dis}.

\section{Phenomenological Description of Quantum Gravitational Decoherence in the Neutrino System}
\label{sec:deco_model}

Modeling possible effects of interactions of any quantum system with quantum
gravitational \ac{dof} without specifying the underlying theory of \ac{qg} becomes
possible in the framework of open quantum systems~\cite{Rivas:2012ugu, Ellis:1983jz, Ellis:1992dz, Huet:1994kr, Chang:1998ea}.
In that framework, only the dynamics of a subset, $S$, of all \ac{dof} is actively considered and the
effects due to interactions of this subsystem with the remaining \ac{dof} (usually referred to as
\textit{the environment}) are taken into account by introducing unitarity violating terms
into the evolution equation of the system.
This way the impurity of the state may increase during propagation which is known as decoherence.
Therefore, open quantum systems are described using the density operator \(\varrho\)
fully characterising the state of the (sub-)system while also allowing for the description of mixed states.

In the following, the \ac{dof} of interest are the neutrino and dark flavors while the environment
includes all other \ac{dof}, like for example the matter surrounding the neutrino path as well as
hypothetical quantum gravitational \ac{dof}.

Our working hypothesis for these quantum gravitational \ac{dof} can be described as
a generalized No-Hair theorem~\cite{Hawking:1975vcx,Page:1980qm}:
We assume that in interactions with microscopic black holes from the \ac{qg} vacuum
only unbroken gauge quantum numbers, i.e. color and electric charge, four momentum\footnote{
  As spacetime may not be perfectly continuously homogeneous even at very small scales
  in theories of quantum gravity due to possible pixellation of spacetime,
  the requirement of four momentum conservation and Lorentz invariance could be relaxed since the
  theory would not be Poincaré symmetric anymore.
} as well as total angular momentum is conserved while global quantum numbers,
such as lepton (family) number, are maximally violated.

This feature makes it necessary to extend the subsystem of neutrino flavors with potentially existing
dark fermionic \ac{dof} sharing the same unbroken gauge quantum numbers as neutrinos.
Due to the assumed \ac{qg} behaviour an initially pure neutrino state could then evolve into a mixed state containing
non-zero contributions from these dark sectors.
This effect could, in principle, be detectable at neutrino telescopes even with atmospheric neutrinos.
In this work we analyze and quantify the potential sensitivity of these experiments on constraining \ac{qg}
model parameters with atmospheric neutrinos.
To this end, we extend the phenomenological description of this effect from the vacuum case,
already extensively discussed in~\cite{Hellmann:2021jyz, Hellmann:2022cgt},
to incorporate all matter effects necessary for a proper description of the propagation of atmospheric neutrinos.

Assuming a memory-less, i.e. Markovian, time evolution, the dynamics of a general open quantum system, $S$, are governed by
the Lindblad equation~\cite{Lindblad:1975ef}:
\begin{align}
  \dot{\varrho}_S &= -i [H_S, \varrho_S] + D[\varrho_S] \,. \label{eq:lindblad}
\end{align}
The first term describes the unitary part of the time evolution due to the system Hamiltonian $H_S$
as in the usual von Neumann equation for the density operator.
The second term, called the dissipator, $D[\varrho_S]$, encodes the effective impact of the environmental degrees
of freedom on the open system $S$ and is responsible for the decoherence effect.

Since in the following we consider (ultra relativistic) atmospheric neutrinos propagating through Earth,
the Hamiltonian\footnote{
  Here we have already subtracted a part proportional to the identity from the Hamiltonian
  which is allowed since only the commutator of $H_S$ appears in the Lindblad equation.
} in the vacuum mass basis is given by
\begin{align}
  H_\nu &= H_0 + H_{\mathrm{matter}}\,,
  &&\mathrm{with}
  &&H_0 = \frac{\Delta M^2}{2 E}\,,
  &&\mathrm{and}
  &&H_{\mathrm{matter}} = U^\dagger V U \,, \label{eq:hamiltonian}
\end{align}
where the mass squared difference matrix is given by $\Delta M^2 = \mathrm{diag}(0, \Delta m_{21}^2, \ldots, \Delta m_{n_{f}1}^2)$
with \(\Delta m_{kj}^2 := m_k^2 - m_j^2\).
The number of fermions, $n_f$, accounts for the number of active neutrino generations as well as for all dark sectors
sharing the same unbroken gauge quantum numbers as the active neutrinos.
As we always consider three active generations $n_f \geq 3$ holds.
By choosing the Hamiltonian~\eqref{eq:hamiltonian}, we implicitly assume that also all other fermions
are ultra relativistic at the energy scales of atmospheric neutrinos.
This assumption is reasonable assuming that \ac{qg} interactions approximately conserve the four momentum
and hence also the center of mass energy of the propagating system.
Therefore only dark sectors with masses within the mass uncertainty of the propagating neutrino wave packets
are allowed to appear in the beam.
Consequently also the additional particles must have masses on the same order of magnitude of the neutrino mass
and would propagate at ultra relativistic velocities.
If the underlying theory of \ac{qg} violates energy momentum conservation at high energy scales this reasoning
breaks down and also particles with masses outside the mass uncertainty of the neutrino beam can be produced.
In this case, it is not guaranteed that the particles appearing in the beam would be ultra relativistic and the corresponding
vacuum Hamiltonian entries become 
\begin{align}
  \langle \psi_{j} \vert H_0 \vert \psi_{j} \rangle = \sqrt{\vec{p}^2 + m_{\psi_{j}}^2} + \delta E_{\mathrm{QG}} \,, \label{eq:LIV}
\end{align}
where $\delta E_{\mathrm{QG}}$ is some potentially Lorentz violating term in the dispersion relation.
In the following, we are only interested in flavor transition rates between active neutrino initial and final states.
As we assume that mixing between active neutrinos and the dark sector is negligible,
the structure of the Lindblad equation~\eqref{eq:lindblad} ensures that these active-active transition rates are independent of
the Hamiltonian components in the dark sector.
% This, of course, is only true as long as the dissipator in equation~\eqref{eq:lindblad} does not give rise to coherence between dark and active \ac{dof},
% which is ensured by the assumption that the \ac{qg} effect selects a certain mass eigenstate when acting on a particle from the ensemble.
% Consequently quantum coherence is erased after such an interaction and the dissipator describing the resulting mean effect on the full ensemble
% shares this coherence damping property.
Therefore, the system Hamiltonian can be assumed to be of the form~\eqref{eq:hamiltonian}.

Moreover, the matter potential $V$ in the flavor basis is given by the charged and neutral current neutrino
forward scattering potentials in ordinary matter, i.e.
\begin{align}
  V &= \mathrm{diag}(V_{\mathrm{CC}} + V_{\mathrm{NC}}, V_{\mathrm{NC}}, V_{\mathrm{NC}}, 0 \dots, 0) \,, 
\end{align}
and it can be translated to the mass basis using the full mixing matrix of the system
\begin{align}
  U := U_{\mathrm{PMNS}} \oplus \mathbb{I}_{(n_{f} - 3) \times (n_{f} - 3)} \,,
\end{align}
where $U_{\mathrm{PMNS}}$ is the $3 \times 3$ \ac{pmns} matrix and we assume that there is little
to no a priori mixing of active neutrinos and the dark sector.
Of course this framework also applies in case there is non-negligible mixing between these states.

Finally, the dissipator contains several terms due to incoherent scattering of high energy neutrinos with the surrounding matter such as neutrino loss and $\nu_\tau$ regeneration as described in~\cite{Arguelles:2021twb}, i.e.
\begin{align}
  D[\varrho_\nu] = D_{\mathrm{matter}}[\varrho_\nu] + D_{\mathrm{QG}}[\varrho_\nu] \,.
\end{align}
Since the matter part of the dissipator is already extensively discussed in the literature~\cite{Gonzalez-Garcia:2005ryx,Arguelles:2021twb},
we want to focus on the details of \(D_{\mathrm{QG}}[\varrho_\nu]\) from now on.
The simplest dissipator respecting the properties of the \ac{qg} interaction discussed above reads:
\begin{align}
  D_{\mathrm{QG}}[\varrho_\nu] &= \sum_{i, j = 0}^{n_{f}^2 - 1} {\mathcal{D}_{\mathrm{QG}}}^{i}{}_{j} \rho^j \lambda_{i}
  &&\mathrm{with}
  &&\mathcal{D}_{\mathrm{QG}} = \mathrm{diag}(0, \Gamma, \ldots, \Gamma) \,,
  \label{eqn:Diss}
\end{align}
where we introduce the usual \(\mathrm{SU}(n_{f})\) generator $\lambda_i$ representation
for the density matrix $\varrho_{\nu} =: \sum_{i}\rho^i \lambda_i$ and the dissipator.
For more details on the choice of basis see appendix A in Ref.~\cite{Hellmann:2021jyz}.

Setting \({\mathcal{D}_{\mathrm{QG}}}^{0}{}_{j}\) and \({\mathcal{D}_{\mathrm{QG}}}^{j}{}_{0}\) to zero means that the \ac{qg} effect
does not lead to additional particle loss or creation.
Only conversions between different fermion types are permitted.
All dissipator elements corresponding to off-diagonal \(\mathrm{SU}(n_{f})\) generators
damp oscillations while those corresponding to the other diagonal \(\mathrm{SU}(n_{f})\)
lead to a uniform distribution of the probability across all fermion generations.
The oscillation damping effect occurs since \ac{qg} interactions are assumed to select a certain mass eigenstate
while the uniform distribution of probability is required by the maximal flavor violation, i.e. the extended No-Hair theorem discussed above.

The exact expression of the damping factor $\Gamma$ is unknown since it depends on the details of the underlying theory of \ac{qg}.
However, we can still employ a phenomenological parametrization in terms of the neutrino energy $E_{\nu}$,
and consider a representative set of models characterized by how $\Gamma$ depends on $E_\nu$:
\begin{equation}
    \Gamma(E_\nu) = \gamma_0 \left(\frac{E_\nu}{E_0}\right)^n \,.
    \label{eqn:Gamma}
\end{equation}
Here, $n$ is the exponent of the energy, $E_0$ is a Ref. energy, and $\gamma_0$ quantifies the strength of the decoherence effect at $E_0$.
This power-law dependence has been explored in the literature~\cite{Ellis:1983jz, Ellis:1992dz, Huet:1994kr,Chang:1998ea,Farzan:2008zv,Carrasco:2018sca,Stuttard:2020qfv,DeRomeri:2023dht}
and considered in experimental searches~\cite{Lisi:2000zt,Fogli:2007tx,Coloma:2018idr,Gomes:2020muc}.
Since \ac{qg} effects are expected to grow stronger as the wave length of the particles approach the Planck length,
it is reasonable to investigate positive values of $n$ (specifically in the case $n = 0$, $1$, $2$), but there are also interesting cases
where a negative $n$ dependence arises, see e.g. Ref.~\cite{Domi:2024ypm}.
In accordance with the literature, we set $E_0 = \SI{1}{\tera\electronvolt}$, to be close to the peak of the neutrino energy distribution
observed in the IceCube experiment.

According to Ref.~\cite{Hellmann:2022cgt} the neutrino oscillation probability in vacuum reads
\begin{align}
    \begin{split}
      P(\nu_\alpha \rightarrow \nu_\beta) &= \frac{1}{n_f} \left(1 - \exp{\left(-\Gamma(E_\nu) L\right)}\right)
      + \left(\sum_{k = 1}^{3} \vert U_{\alpha k}\vert^2 \vert U_{\beta k}\vert^2\right) \exp{\left(-\Gamma(E_\nu) L\right)} \\
      &\quad + 2 \sum_{j > i = 1}^{3} \mathrm{Re}\left[U_{\alpha j}^{\ast} U_{\alpha i} U_{\beta j} U_{\beta i}^{\ast}\right] \exp{\left(-\Gamma(E_\nu) L -\left(\frac{L}{L_{ij}^\text{coh}}\right)^2\right)} \cos\left(\frac{\Delta m_{ij}^2 L}{2 E_\nu}\right) \\
      &\quad- 2 \sum_{j > i = 1}^{3} \mathrm{Im}\left[U_{\alpha j}^{\ast} U_{\alpha i} U_{\beta j} U_{\beta i}^{\ast}\right] \exp{\left(-\Gamma(E_\nu) L -\left(\frac{L}{L_{ij}^\text{coh}}\right)^2\right)} \sin\left(\frac{\Delta m_{ij}^2 L}{2 E_\nu}\right)
    \end{split} \label{eq:oscprob}
\end{align}
assuming the dissipator from equation~\eqref{eqn:Diss} with the baseline $L$ and $\Gamma$ as given in equation~\eqref{eqn:Gamma}.
For the cases $n \neq 0$, it is convenient to define the so called coherence energy $E_{\mathrm{QG}}$, where
\begin{align}
    \Gamma(E_{\mathrm{QG}}) L \stackrel{!}{=} 1 &&\Leftrightarrow &&E_{\mathrm{QG}} = E_0 \left(\frac{1}{\gamma_0 L}\right)^{\frac{1}{n}} \,, \label{eq:Eqg}
\end{align}
indicating the neutrino energy at which the \ac{qg} effect becomes dominant.
Furthermore, we observe that the decoherence effect depends exponentially on the baseline $L$, which is why the strongest impact is expected for astrophysical neutrinos,
as pointed out in~\cite{Klapdor-Kleingrothaus:2000kdx}.
In the context of astrophysical neutrinos, also the \ac{wp} separation effect~\cite{
  Kayser:1981ye,Frampton:1982qi,Giunti:1991ca,Giunti:1993se,Giunti:1997sk,
  Kiers:1997pe,Akhmedov:2009rb,Akhmedov:2014ssa,Kersten:2015kio,Naumov:2020yyv} becomes significant as the travel distances involved may exceed the coherence length,
\begin{align}
  L_{ij}^{\mathrm{coh}} &= 4\sqrt{2} \sigma_x \frac{E_\nu^2}{\Delta m_{ij}^2} \,, \label{eq:L_coh}
\end{align}
over which the neutrino \acp{wp} overlap with each other and coherence is maintained.
Here $\sigma_x$ is the average position space \ac{wp} width of the neutrino mass eigenstates.
This introduces an additional coherence damping exponential\footnote{Here we assume Gaussian shaped neutrino wave packets.}
$\exp(-(L / L_{ij}^{\mathrm{coh}})^2)$ as a multiplicative factor in the oscillatory terms in equation~\eqref{eq:oscprob}.
Note that \ac{wp} separation represents only a subleading effect for atmospheric neutrinos and is, hence, neglected in the respective parts of this analysis.

\begin{figure}[!t]
  \centering
\includegraphics{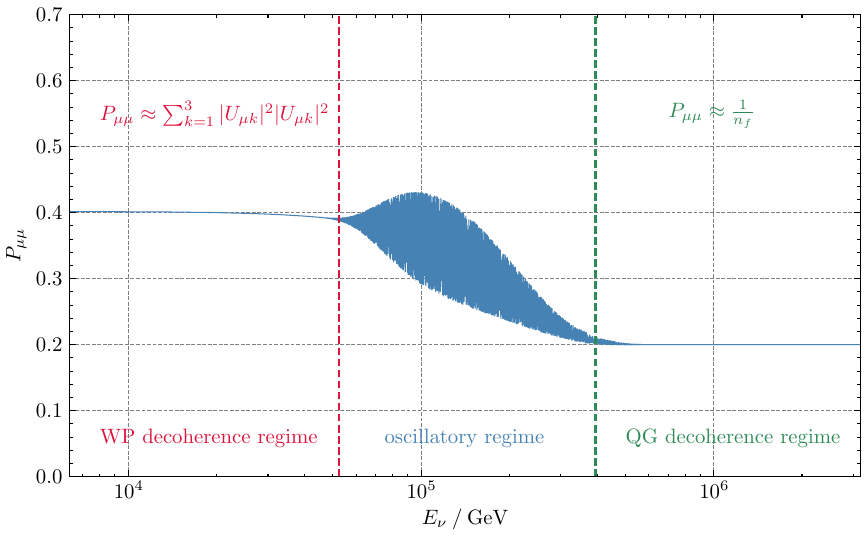}
  \caption{Oscillation probability of $\nu_\mu \rightarrow \nu_\mu$ for the baseline $L = 14.4 \, \mathrm{Mpc}$ (the distance to NGC 1068),
  a wave packet width $\sigma_x = 10^{-9} \, \mathrm{m}$, the \ac{qg} parameter $\gamma_0 = 10^{-35} \, \mathrm{eV}$,
  the fermion number of $n_f = 5$ and the quadratic decoherence model, i.e. $n = 2$.
  The mass squared differences and mixing parameters are extracted from nuFit v5.3~\cite{Esteban:2020cvm, nuFit:v53}
  and rounded to the first digit that is not subject to uncertainties.}
  \label{fig:oscprob_asymp}
\end{figure}

Figure~\ref{fig:oscprob_asymp} shows the $\nu_\mu \to \nu_\mu$ oscillation probability for the exemplary distance of $L = 14.4 \, \mathrm{Mpc}$
to demonstrate the different asymptotic behaviors of the vacuum oscillation probability of astrophysical neutrinos:
At low energies the group velocities of the neutrinos are sufficiently different such that the \acp{wp} are
out of contact by the time they arrive at the detector leading to a constant oscillation probability in the \ac{wp} decoherence limit.
At very high energies the \ac{qg} decoherence effect dominates damping the oscillation probabilities to $1 / n_f$ as discussed above.
It is worth noting that in the \ac{qg} limit the final flavor ratios of active neutrinos will always be democratic, i.e. $(1:1:1)$,
regardless of the flavor composition at the source.

Depending on the \ac{qg} and \ac{wp} parameters there may be also a coherent, oscillatory regime in between the two asymptotic cases.
The respective oscillations of the flavor transition probabilities would in principle be observable at neutrino telescopes
if the energy resolution is sufficiently high, i.e.
\begin{align}
  \frac{\Delta E_{\nu}}{E_{\nu}} \lesssim \frac{2 E_{\nu}}{\Delta m^2 L} \,,
\end{align}
where $\Delta E_{\nu} / E_{\nu}$ is the relative energy resolution of the detector, $\Delta m^2$
is the mass squared splitting corresponding to the oscillation to be resolved
and $L$ is the distance between neutrino source and detector.

Despite the superior potential sensitivity of astrophysical neutrinos for \ac{qg}-induced decoherence the limited statistics of present data
makes it diffcult to extract reliable constraints (see the discussion in section~\ref{ssec:orca}).
As a result, atmospheric neutrinos currently represent the most suitable candidates for conducting \ac{qg} searches with positive energy dependence.

\section{Sensitivity Estimation for the Decoherence Parameter Across All Models}
\label{sec:analysis}
In this section, we assess the potential of the IceCube Neutrino Observatory to detect \ac{qg}-induced quantum decoherence in the presence of new fermionic particles.
The analysis is based on the publicly available \ac{mc} simulation sample used in Ref.~\cite{ICECUBE:2024fej} and takes into account all systematic effects discussed in Refs.~\cite{IceCube:2020tka,ICECUBE:2024fej}, with no real data used in this work.
Firstly, we explain in section~\ref{ssec:mc} details of the employed \ac{mc} sample and how we derive the theory prediction of the various models, before detailing 
the statistical approach in section~\ref{ssec:stat}.
In section~\ref{ssec:icecube}, we present the expected sensitivities of the IceCube experiment 
to \ac{qg} parameters and new particles. Finally, in section~\ref{ssec:orca} we discuss potential future analyses aimed at optimising the sensitivity to these parameters.

\subsection{Propagation of Fluxes and Detector Response Modelling}
\label{ssec:mc}
To test the model presented in Ref. \cite{Hellmann:2021jyz, Hellmann:2022cgt}, we focus on high-energy atmospheric neutrinos. In this energy regime, the relevant processes affecting neutrino propagation through the Earth are driven by incoherent scattering of neutrinos with the Earth matter, leading to neutrino absorption~\cite{Cooper-Sarkar:2011jtt}, energy degradation and tau neutrino regeneration~\cite{Beacom:2001xn, Arguelles:2022bma}. In order to incorporate all these effects for a given initial neutrino flux, we use a modified version of the NuSQuIDS~\cite{Arguelles:2021twb} public software, in which we have incorporated an additional decoherence term to account for the potential presence of dark fermionic particles, as described in section~\ref{sec:deco_model}. As an initial conventional and prompt atmospheric neutrino flux, we use the nuflux package from IceCube~\cite{IceCube_Collaboration2023-sd} for a H3a\_SIBYLL23C model. The astrophysical flux is modelled by a single unbroken power law in neutrino energy with a spectral index of \num{-2.5} \cite{IceCube:2020acn_power_law}. The resulting final flux is then determined for different numbers of fermions $n_f$ and the different decoherence models. To incorporate the dependence of the decoherence term on the \ac{qg} parameter $\gamma_0$, the final flux is computed and then interpolated across the corresponding $\gamma_0$ values.

To convert the resulting final fluxes into actual events at IceCube, we use the public MEOWS \ac{mc} sample~\cite{DVN/9WGYQN_2024} and associated tools. These facilitate the transformation from the predicted neutrino flux to detected events, following a similar approach to that described in Ref.~\cite{ICECUBE:2024fej}. This Monte Carlo release corresponds to $T = 7.6$ years of livetime. The \ac{mc} sample contains \num{24902627} unweighted events, which correspond to \num{305735} neutrino events weighted within the standard oscillation scenario. The released MEOWS MC dataset consists of only track-like neutrino events, i.e. $\nu_\mu$ and $\bar{\nu}_\mu$, whose available information are the true and reconstructed energy and cosine of the zenith angle. The reconstruction achieves an energy resolution of $\sigma_{\text{log}_{10}(E_{\nu_\mu})} \sim 0.3$ and an angular resolution $\sigma_{\cos{\theta_\text{zenith}}}$ between 0.005 and 0.015 as a function of the energy~\cite{IceCube:2020tka}. 

In order to incorporate the new decoherence effects we reweight the \ac{mc} events with the appropriate oscillation probabilities evaluated with NuSQuIDS.
We also take into account the systematic effects of the detector and the initial flux models by allowing for an additional, dynamical rescaling of the MC weights depending on the set of nuisance parameters $\vec{\eta}$ described in Refs.~\cite{IceCube:2020tka,ICECUBE:2024fej} and listed in table~\ref{tab:nuisance}.
\begin{table}[]
  \caption{A list of all nuisance parameters, $\eta_j$, their nominal values, standard deviations and box constraints taken into account in this analysis.
  For more details see Refs.~\cite{IceCube:2020tka,ICECUBE:2024fej}.}
  \begin{center}
  \begin{tabular}{c c c c} 
  \toprule
  Parameter $\eta_j$ & Nominal value $\bar{\eta}_j$ & Standard Deviation $\sigma_{\eta_j}$ & Box constraint\\
  \midrule
  DOM efficiency & $\num{0.97}$ & $\num{0.1}$ & $[\num{0.94},\num{1.03}]$\\
  Bulk Ice Gradient 0 & $\num{0.0}$ & $\num{1.0}$ & NA\\
  Bulk Ice Gradient 1 & $\num{0.0}$ & $\num{1.0}$ & NA\\
  Forward Hole Ice $(p_2)$ & $\num{-1.0}$ & $\num{10.0}$ & $[\num{-5},\num{3}]$\\
  Normalization $(\Phi_\text{conv.})$ & $\num{1.0}$ & $\num{0.4}$ & NA\\
  Spectral shift $(\Delta \gamma_\text{conv.})$ & $\num{0.00}$ & $\num{0.03}$ & NA\\
  Atm. Density & $\num{0.0}$ & $\num{1.0}$ & NA\\
  Barr WM & $\num{0.0}$ & $\num{0.40}$ & $[\num{-0.5},\num{0.5}]$\\
  Barr WP & $\num{0.0}$ & $\num{0.40}$ & $[\num{-0.5},\num{0.5}]$\\
  Barr YM & $\num{0.0}$ & $\num{0.30}$ & $[\num{-0.5},\num{0.5}]$\\
  Barr YP & $\num{0.0}$ & $\num{0.30}$ & $[\num{-0.5},\num{0.5}]$\\
  Barr ZM & $\num{0.0}$ & $\num{0.12}$ & $[\num{-0.25},\num{0.5}]$\\
  Barr ZP & $\num{0.0}$ & $\num{0.12}$ & $[\num{-0.2},\num{0.5}]$\\
  Normalization $(\Phi_\text{astro})$ & $\num{0.787}$ & $\num{0.36}$ & NA\\
  Spectral shift $(\Delta \gamma_\text{astro})$ & $\num{0.0}$ & $\num{0.36}$ & NA\\
  Cross section $\sigma_{\nu_\mu}$ & $\num{1.00}$ & $\num{0.03}$ & $[\num{0.5},\num{1.5}]$\\
  Cross section $\sigma_{\bar{\mu}_{\mu}}$ & $\num{1.000}$ & $\num{0.075}$ & $[\num{0.5},\num{1.5}]$\\
  Kaon energy loss $\sigma_{KA}$ & $\num{0.0}$ & $\num{1.0}$ & NA\\
  \bottomrule
  \end{tabular}
  \label{tab:nuisance}
  \end{center}
\end{table}

\subsection{Statistical Approach}
\label{ssec:stat}
To estimate the constraints that neutrino oscillation experiments can place on the \ac{qg}-parameter $\gamma_0$,
we conduct a likelihood ratio test on the binned, reweighted MC event counts.
In order to correctly take into account all relevant systematic effects as well as \ac{mc} modeling uncertainties in the analysis of the counting experiment under consideration, we employ a profile likelihood function that is given by a product of a likelihood describing the bin counts, $\mathcal{L}_{\text{bin}}$, and a penalty term, $\Pi(\vec{\eta})$, encoding the distribution information of the nuisance parameters $\vec{\eta}$.
For a given data set, $\vec{X}$, and a chosen decoherence model, i.e. for fixed $n_f$ and $n$, the profile likelihood function is then obtained by maximizing over the set of nuisance parameters:
\begin{equation}
    \mathcal{L}_{\text{profile}} (\gamma_0 \vert \vec{X}) := \sup_{\vec{\eta}} \mathcal{L}_{\text{bin}}({\gamma_0, \vec{\eta}} \vert \vec{X}) \Pi(\vec{\eta}) \,.
\end{equation}
We model the nuisance penalty term,
\begin{equation}
    \Pi (\vec{\eta}) = \prod_{j=1}^{N_\text{sys}} \frac{e^{- \frac{(\eta_j - \bar{\eta}_{j})^2}{2\sigma_{\eta_j}^2}}}{\sqrt{2 \pi \sigma_{\eta_j}^2}} \,,
\end{equation}
as a product of independent, one-dimensional normal distributions, with nominal values, $\bar{\eta}_j$, and standard deviations, $\sigma_{\eta_j}$, taken from reference~\cite{IceCube:2020tka} and specified in table~\ref{tab:nuisance}.

The likelihood function describing the distribution of bin counts is taken as~\cite{Arguelles:2019izp},
\begin{align}
  \mathcal{L}(\gamma_0, \vec{\eta} \vert \vec{X}) &= \prod_{l = 1}^{n_{\mathrm{bins}}} \mathcal{L}_\text{eff}(n_{\mu, l}(\gamma_0, \vec{\eta}), \sigma_l(\gamma_0, \vec{\eta}) \vert X_l)\,,
\end{align}
where $n_{\mathrm{bins}}$ is the total number of bins and $X_l$ is the measured number of events in bin $l$.
Furthermore, the expected number of event counts, $n_{\mu,l}(\gamma_0, \vec{\eta})$, and the associated \ac{mc} uncertainty, $\sigma_{\text{MC}, l}^2(\gamma_0, \vec{\eta})$, in bin $l$ predicted according to a given decoherence model are computed from the \ac{mc} weights, $\omega_{i}^l$, as follows:
\begin{align}
    n_{\mu, l} (\gamma_0, \vec{\eta}) &= \sum_{i = 1}^{N_l} \omega_i^l(\gamma_0, \vec{\eta})\,,
    &\sigma_{\text{MC}, l}^2(\gamma_0, \vec{\eta}) &= \sum_{i = 1}^{N_l} \left[\omega_{i}^l (\gamma_0, \vec{\eta})\right]^2\,.
\end{align}
Here, the $i$-summation runs over all weights associated to the $l$-th bin.
The bin-wise likelihood function $\mathcal{L}_\text{eff}$ is an adaption of the Poisson likelihood, specifically designed to account for statistical uncertainties from MC simulations, first introduced in reference~\cite{Arguelles:2019izp}.

The logarithmic likelihood ratio test statistic is then defined as
\begin{align}
  \ln\Lambda &:= -2\ln\left(\frac{\sup_{\gamma_0 \in \Theta_0} \mathcal{L}_{\text{profile}}(\gamma_0 \vert \vec{X})}{\sup_{\gamma_0 \in \Theta} \mathcal{L}_{\text{profile}}(\gamma_0 \vert \vec{X})}\right) \,, \label{eq:likelihood}
\end{align}
where the parameter space of the null hypothesis, $\Theta_0 = \{\gamma_0^0\}$, consists of a single parameter configuration that we wish to test.
As the toy data set, we take the prediction of the standard case, i.e. $X_{i} = n_{\mu, i}^{\mathrm{std}} := n_{\mu, i}(\gamma_0 = 0, \vec{\eta})$.
For $\gamma_0 = 0$ the event counts in each bin are independent of the number of fermions $n_{f}$ and energy dependence $n$ of the decoherence model.
The full parameter space, $\Theta$, for each combination $(n_{f}, n)$ in the null hypothesis can be taken as $\Theta = \mathbb{R}$.
Thus, for a given choice $(n_{f}, n)$ we perform a one dimensional likelihood ratio test for the \ac{qg} parameter $\gamma_0$.

In the limit of large event counts, the log likelihood ratio converges to the following, simple $\chi^2$ test statistic:
\begin{equation}
  \ln\Lambda(\gamma_0) \to \chi^2(\gamma_0) = \min_{\vec{\eta}} \sum_{l = 1}^{n_\text{bins}} \frac{(n_{\mu,l}^\text{std} - n_{\mu, l}(\gamma_0, \vec{\eta}))^2}{n_{\mu, l}(\gamma_0, \vec{\eta}) + \sigma_{\text{MC}, l}^2 (\gamma_0, \vec{\eta})} + 
  \sum_{j = 1}^{N_{\text{sys}}} \frac{(\eta_j - \bar{\eta}_j)^2}{\sigma_{\eta_j}^2}\,.
\end{equation}
According to Wilk's theorem $\ln(\Lambda)$ follows a $\chi^2$ distribution with one DOF
since $\Theta_0 \subset \mathrm{interior}(\Theta)$ and the large event count limit applies to good approximation.
This allows us to specify the \acp{cl} of the $\gamma_0$ exclusion limits by 
comparing the respective $\chi^2(\gamma_0)$ value to the threshold values of the $\chi^2$ distribution. 
%Since this work aims to demonstrate the potential for such an analysis, we do not account for any systematic uncertainties. This approach can inspire future research involving a comprehensive analysis with experimental data.

\subsection{Expected Sensitivities for the IceCube Neutrino Observatory}
\label{ssec:icecube}
\begin{figure}[!t]
  \centering
  \includegraphics{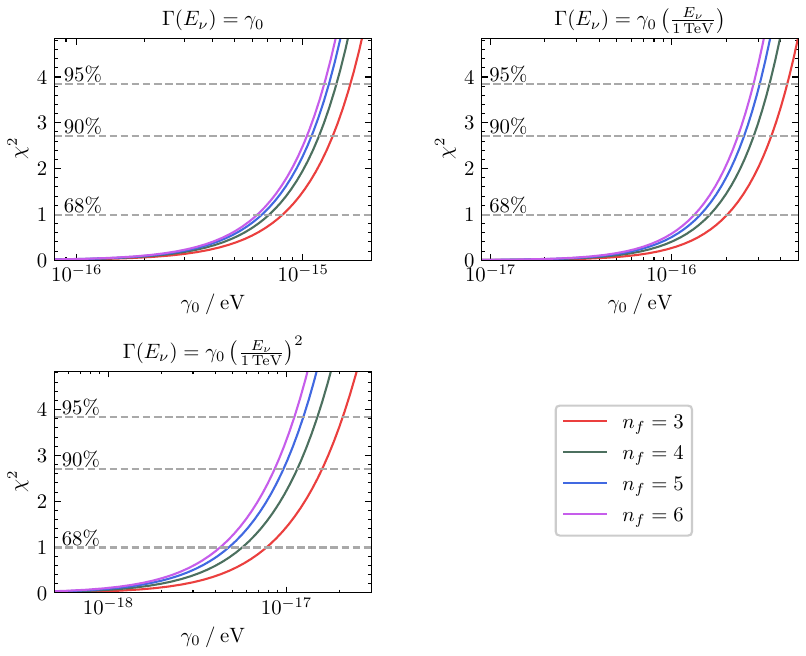}
  \caption{Estimated IceCube sensitivity for $T=7.6$ years of livetime, for the three decoherence models under consideration, each with varying fermion numbers. The $\chi^2$ threshold values for the 68\%, 90\% and 95\% \acp{cl} are also reported as dashed lines.}
  \label{fig:chi2}
\end{figure}

\begin{figure}[!t]
  \centering
  \includegraphics[width = 0.5\textwidth]{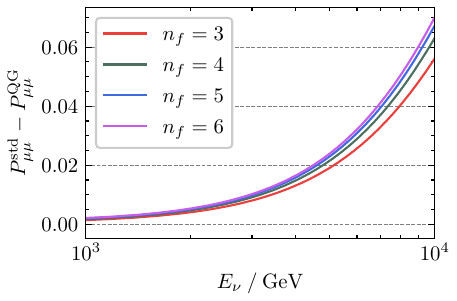}
  \caption{Difference between the standard and \ac{qg} model prediction of the $\nu_{\mu} \to \nu_{\mu}$ transition probability,
  $P_{\mu\mu}^\text{std}$ and $P_{\mu\mu}^\text{QG}$, for various numbers of fermions in the quadratic decoherence model,
  with $\cos{\theta_\text{zenith}} = -1$, corresponding to a baseline equal to the Earth diameter.
  Here, the model probabilities are evaluated at the $95\%$ \ac{cl} $\gamma_0 = \SI{2.08e-17}{\eV}$ for $n_f = 3$ and $n = 2$.}
  \label{fig:1dosci}
\end{figure}

In this section, we determine the expected potential of the IceCube experiment to probe the \ac{qg} parameters presented in this work.
Figure~\ref{fig:chi2} presents the dependence of $\chi^2$ on the decoherence parameter $\gamma_0$ for the various decoherence models discussed in section~\ref{sec:deco_model}, differing in their energy-dependence, $n$, and in the number of fermions, $n_f$.
The 68\%, 90\% and 95\% \acp{cl} are indicated by dashed lines.
Table~\ref{tab:chi2_values} reports the 90\% \ac{cl} values for the \ac{qg} parameter $\gamma_0$ for the different energy dependencies which are,
for $n_f = 3$, in good agreement with the literature~\cite{ICECUBE:2024fej}.
\begin{table}[!t]
  \caption{The $\gamma_0 / \text{eV}$ values at 90\% \ac{cl} for the different energy dependencies and $n_f \in \{3, 4, 5, 6\}$.
  In parentheses we report the idealized bounds obtained in a statistics-only approach, i.e. neglecting systematic effects.}
  \begin{center}
  \begin{tabular}{c c c c} 
  \toprule
  \multirow{2}{*}{$n_f$} & \multicolumn{3}{c}{$\Gamma(E_\nu)$}\\
  \cmidrule(l){2-4} 
  & $\gamma_0$ & $\gamma_0 \left(\frac{E_\nu}{1 \text{TeV}}\right)$ &  $\gamma_0 \left(\frac{E_\nu}{1 \text{TeV}}\right)^2$\\
  \midrule
  3 & $\num{1.35e-15}$ ($\num{1.52e-16}$) & $\num{3.54e-16}$ ($\num{5.20e-17}$) & $\num{1.60e-17}$ ($\num{2.76e-18}$)\\
  4 & $\num{1.18e-15}$ ($\num{1.33e-16}$) & $\num{2.83e-16}$ ($\num{4.37e-17}$) & $\num{1.16e-17}$ ($\num{2.11e-18}$)\\
  5 & $\num{1.09e-15}$ ($\num{1.24e-16}$) & $\num{2.51e-16}$ ($\num{3.99e-17}$) & $\num{9.68e-18}$ ($\num{1.81e-18}$)\\
  6 & $\num{1.04e-15}$ ($\num{1.18e-16}$) & $\num{2.32e-16}$ ($\num{3.76e-17}$) & $\num{8.63e-18}$ ($\num{1.65e-18}$)\\
  \bottomrule
  \end{tabular}
  \label{tab:chi2_values}
  \end{center}
\end{table}
In addition to the bounds obtained taking into account systematic effects of the experiment, we also show the equivalent bounds derived from a statistics-only approach in Table~\ref{tab:chi2_values}.
This comparison provides an estimate of the impact of systematic uncertainties.
The analysis shows that the sensitivity to the \ac{qg} parameter strengthens, as the number of fermions increases, which aligns with expectations.
This behavior is anticipated because the decoherence limit diverges from the standard case with an increasing number of fermions, as illustrated in Figure~\ref{fig:1dosci}.
Moreover, the bounds become more stringent as the power of the energy dependence in the decoherence model increases.
Considering Figure~\ref{fig:1dosci}, we see that for an energy-dependent decoherence model the largest deviation is expected in high-energy neutrinos.
At the $95\%$ \ac{cl} bounds determined in this analysis the difference to the standard $\nu_{\mu} \to \nu_{\mu}$ probability is about $6\%$ at $E_{\nu} \sim 10\,\mathrm{TeV}$.

\begin{figure}[!t]
  \centering
  \includegraphics[width=\textwidth]{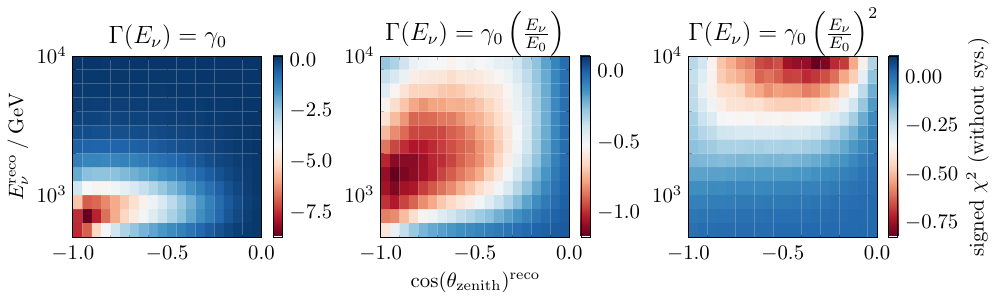}
  \caption{Signed $\chi^2$ values (without systematic effects) for $n_\text{f} = 4$ and the three energy dependencies as a function of the reconstructed zenith angle $\cos{(\theta_\text{zenith})}^\text{reco}$ and reconstructed energy $E_\nu^\text{reco}$, corresponding to the $95\%$ \ac{cl} $\gamma_0$ values: $\gamma_0 \in \{\num{1.41e-15}, \num{3.46e-16}, \num{1.50e-17}\}\,\mathrm{eV}$ (from left to right).}
  \label{fig:signedchi2_wo_sys}
\end{figure}

\begin{figure}[!t]
\centering
  \includegraphics[width = \textwidth]{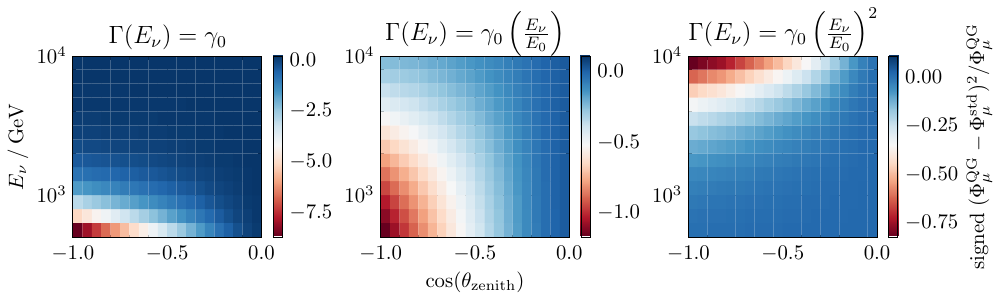}
  \caption{Cross-checks for the signed $\chi^2$ distributions ($n_f = 4$) in terms of the binned, true $\nu_{\mu}$ fluxes in the energy $E_\nu$ and cosine zenith angle $\cos(\theta_\text{zenith})$ plane, scaled such that the minimum of the shown quantity and that of the associated signed $\chi^2$ match.
  Each panel corresponds to a different energy dependence of the underlying \ac{qg} model, i.e. $n \in \{ 0, 1, 2 \}$ (from left to right) and the associated $95\%$ \ac{cl} bounds $\gamma_0 \in \{\num{1.41e-15}, \num{3.46e-16}, \num{1.50e-17}\}\,\mathrm{eV}$.}
  \label{fig:signedchi2_true}
\end{figure}

To further explore the relevant energy and cosine zenith regions that contribute to the IceCube sensitivity, we use the signed $\chi^2$ values per bin defined by
\begin{equation}
    \text{signed} \, \chi_l^2 = \frac{(n^\mathrm{QG}_{\mu,l} - n_{\mu,l}^\text{std}) \vert n^\mathrm{QG}_{\mu,l} - n_{\mu,l}^\text{std}\vert}{n^\mathrm{QG}_{\mu,l}}\,,
\end{equation}
where $n^\mathrm{QG}_{\mu,l}$ are the binned number of events for the respective model with a fixed \ac{qg} parameter $\gamma_0$ and $n_{\mu,i}^\text{std}$ are the binned number of events in the standard case.
Figure~\ref{fig:signedchi2_wo_sys} presents the signed $\chi^2$ computed from the nominal \ac{mc} weights, i.e. without taking into account systematic effects, for $n_f = 4$ and the different decoherence models as a function of reconstructed zenith angle $\cos{\theta_\text{zenith}}^\text{reco}$ and reconstructed energy $E_\nu^\text{reco}$, for $\gamma_0$ such that $\chi^2 = 3.85$, corresponding to the 95\% CL, and the respective $\gamma_0$.  
As a cross-check, Figure~\ref{fig:signedchi2_true} shows an analogous quantity representing the binned true $\nu_{\mu}$ fluxes.
Comparing both Figures we can see that the effect of energy and directional reconstruction on average results in a shift of the regions of highest sensitivity to higher reconstructed energies and cosine zenith values.
From the sensitivity distribution in terms of the true energy and cosine zenith it can be observed that,
in the case of $\Gamma(E) \equiv \gamma_0$, the highest sensitivities are expected in the region of lower cosine zenith, corresponding to up-going events, and lower energies.
The latter is explained by the much higher statistics in the low energy bins compared to the high energy bins.
In contrast, for the other energy-dependent decoherence models, the sensitivity shifts to higher energies as the impact of the \ac{qg} effect increases in the high-energy bins.

\begin{figure}[!t]
  \centering
  \includegraphics[width=\textwidth]{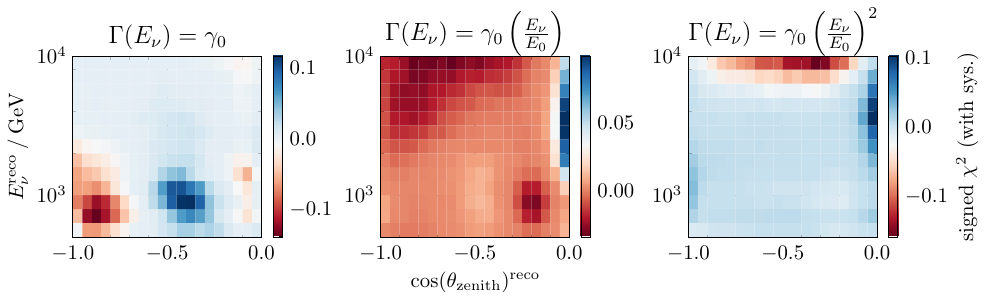}
  \caption{Signed $\chi^2$ values (with systematic effects) for $n_\text{f} = 4$ and the three energy dependencies as a function of the reconstructed zenith angle $\cos{(\theta_\text{zenith})}^\text{reco}$ and reconstructed energy $E_\nu^\text{reco}$, corresponding to the $95\%$ \ac{cl} $\gamma_0$ values: $\gamma_0 \in \{\num{1.41e-15}, \num{3.46e-16}, \num{1.50e-17}\}\,\mathrm{eV}$ (from left to right).}
  \label{fig:signedchi2_w_sys}
\end{figure}

In order to estimate how the systematic effects impact the regions of highest sensitivity, we also compute the signed $\chi^2$ values from the rescaled \ac{mc} events at the corresponding best fit value of the nuisance parameter vector $\vec{\eta}$ obtained from the profile likelihood procedure.
This quantity is shown in Figure~\ref{fig:signedchi2_w_sys} again for $n_f = 4$ and the different energy dependencies as a function of the reconstructed cosine zenith and energy.
From the comparison of the signed $\chi^2$ distributions with and without systematic effects, i.e. Figures~\ref{fig:signedchi2_w_sys} and~\ref{fig:signedchi2_wo_sys}, we observe that the systematic effects can induce significant deviations between the naively expected and the actual regions of highest sensitivity.
Across all decoherence models we see that the overall sensitivity is drastically reduced due to the rescaling of the nominal \ac{mc} weights which is of course expected as we minimize $\chi^2$ on the set of nuisance parameters.
For the energy-independent decoherence model, i.e. $n=0$, we still see that the largest (negative) deviation of $\text{signed}\,\chi^2_l$, caused by the model effects, is located in the up-going, low-energy bins as expected, but we also find a region of positive deviation caused by the application of bin-wise scale factors describing the systematic effects after minimizing $\chi^2$ with respect to the set of nuisance parameters.
In case of the $n=1$ decoherence model, we find that after applying the systematic effects at the best fit point $\vec{\eta} = \vec{\eta}_{\text{opt}}$, the region of maximal sensitivity is further smeared across all energy and cosine zenith bins.
Lastly, for the quadratically energy-dependent decoherence model, as in the case $n = 0$, the region of highest sensitivity remains mostly intact compared to the corresponding region shown in Figure~\ref{fig:signedchi2_wo_sys}.

\subsection{Future prospects}
\label{ssec:orca}
As discussed in section~\ref{sec:deco_model}, astrophysical neutrinos are ideal candidates for this analysis due to their longer baselines and higher energies compared to atmospheric neutrinos. However, this approach is currently impractical due to limited statistics. Ref. \cite{IC_Science_ngc} presents the results of a nine-year (2011–2020) time-integrated analysis of IceCube data, identifying three astrophysical neutrino sources with higher statistical local significance (pre-trial): NGC 1068 with 79 signal events at 5.2$\sigma$, PKS 1424+240 with 77 signal events at 3.7$\sigma$, and TXS 0506+056 with 5 signal events at 3.5$\sigma$. 

\begin{figure}[!t]
\centering
  \begin{subfigure}[b]{0.49\textwidth}
    \includegraphics[width = \textwidth]{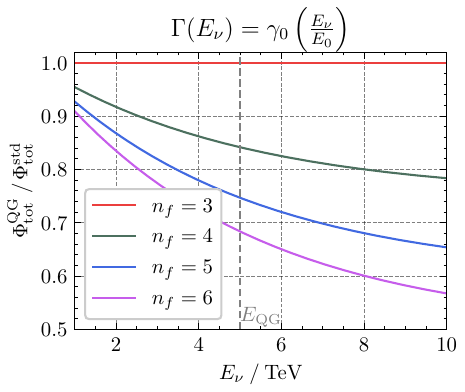}
    %\caption*{}
  \end{subfigure}
  \begin{subfigure}{0.49\textwidth}
    \includegraphics[width = \textwidth]{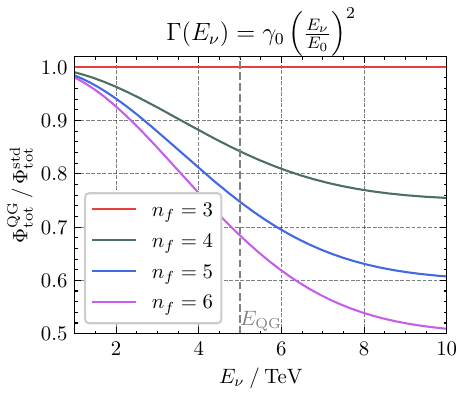}
    %\caption*{}
  \end{subfigure}
  \caption{Total flux $\Phi^\text{QG}_\text{tot}$ normalized to the total standard flux $\Phi^\text{std}_\text{tot}$ for $L = \num{14.4} \, \mathrm{Mpc}$ for the $n = 1$ and $n = 2$ decoherence models in the case of $E_\text{min} < E_\text{QG} = 5 \, \mathrm{TeV} < E_\text{max}$ and for an initial $(1:2:0)$ neutrino flavor ratio.}
  \label{fig:totflux_ngc_Case2}
\end{figure}

Among the cited sources, NGC 1068 is particularly interesting due to the relatively narrow energy range of the detected events,
concentrated in the $\sim[1, 10]$ TeV region.
Figure~\ref{fig:totflux_ngc_Case2} shows the ratio of the expected flux from NGC 1068 in the \ac{qg} scenario, normalized to
the standard flux $\Phi_\text{tot}^\text{std}$, for neutrino energies within the $[1, 10]$ TeV region
and for $E_{\mathrm{QG}} = 5 \,\mathrm{TeV}$, c.f. equation~\eqref{eq:Eqg}.
We model the flux of neutrino flavor $\alpha$ at Earth as $\phi_{\alpha}^{\mathrm{Earth}} := \sum_{\beta} P_{\beta\alpha} \phi_{\beta}^{\mathrm{source}}$
and the total flux as $\phi_{\mathrm{tot}} := \sum_{\alpha} \phi_{\alpha}^{\mathrm{Earth}}$,
where $P_{\beta\alpha}$ is the $\nu_\beta \to \nu_\alpha$ flavor transition probability.
Moreover, we model the source flux\footnote{
    To be precise, the quantity $\phi_{\alpha}^{\mathrm{source}}$ describes the expected flux of neutrino flavor $\nu_\alpha$ at Earth if neutrinos did not oscillate.
    Therefore, it maintains the initial energy spectrum and flavor composition but also takes into account the spreading of the neutrino beam.
} of neutrino flavor $\nu_\alpha$, $\phi_{\alpha}^{\mathrm{source}} := c_{\alpha} \Phi_0 E^{-\sigma}$, as a simple power-law spectrum with spectral index $\sigma$,
common amplitude $\Phi_0$ and initial flavor ratios $c_{\alpha}$ normalized such that $1=\sum_{\alpha}c_{\alpha}$.
The flux ratio, $\Phi_\text{tot}^\text{QG} / \Phi_\text{tot}^\text{std}$, shown in Figure~\ref{fig:totflux_ngc_Case2} is therefore independent of $\Phi_0$, $\sigma$ and the initial flavor ratio $c_{\alpha}$ under these assumptions.
Note that $\Phi_\text{tot}^\text{QG} / \Phi_\text{tot}^\text{std}$ represents a purely theoretical quantity used to emphasize how the \ac{qg} model differs from the standard prediction given the same initial flux spectrum.
This however cannot be used to analyze an actual data set since both the standard and \ac{qg} predictions need to be fitted to the data individually resulting in possibly different parameter values $\Phi_0$, $\sigma$, and so on.

If $E_{QG}$ is located within or close to the interval of observed neutrino energies, a prominent dip towards the asymptotic limit
$\phi_{\mathrm{tot}}^{\mathrm{QG}} \to 3 / n_{f} \phi_{\mathrm{tot}}^{\mathrm{std}}$ can be observed, c.f. Figure~\ref{fig:totflux_ngc_Case2}.
This characteristic could allow for an energy-dependent \ac{qd} analysis.
However, such an analysis would likely need to be model-dependent, assuming the currently fitted spectrum is correct.
For a more realistic and model-independent analysis, where different spectral models, source flavor compositions, and additional quantum decoherence parameters are simultaneously fitted, the current available statistics are insufficient to achieve a statistically significant result.

If $E_{QG} > E_{\mathrm{max}}$, where $E_{\mathrm{min}} / E_{\mathrm{max}}$ are the lower and upper boundaries of the observed energy region,
\ac{qg}-induced decoherence has not yet become dominant.
In this case, the \ac{qg} fluxes approximately match the standard prediction.
The last possible scenario, $E_{QG} < E_{\mathrm{min}}$, corresponds to the case where the system is already in the \ac{qd} limit and the total flux
reached its asymptotic value, $\phi_{\mathrm{tot}}^{\mathrm{QG}} = 3 / n_{f} \phi_{\mathrm{tot}}^{\mathrm{std}}$.
Thus the \ac{qd} effect would only result in a rescaling of the flux normalization and could not be estimated separately by fitting the neutrino fluxes.

The number of events collected per source and the required observation years indicate that it will take a significant amount of time before an astrophysical analysis can place meaningful constraints on these parameters.

In this respect, in the coming years, it may be beneficial to conduct a combined atmospheric neutrino analysis between IceCube and KM3NeT. Such an analysis would not only benefit from the additional statistics provided by KM3NeT but also from the different systematics affecting the analysis, enabling stronger statistical constraints on the quantum decoherence parameters.

\begin{figure}[!t]
  \centering
  \includegraphics{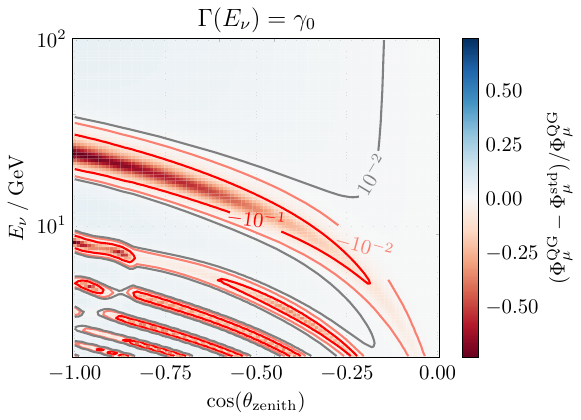}
  \caption{Muon neutrino oscillogram for $n_f = 4$ in case of the constant decoherence model and for the corresponding $95\%$ \ac{cl} $\gamma_0 = \SI{1.41e-15}{\eV}$ value.
  Shown is the relative muon neutrino flux deviation as a function of the energy $E_\nu$ and the zenith angle $\cos{\theta_\text{zenith}}$.}
  \label{fig:orca}
\end{figure}

On the KM3NeT side, ARCA would be the most suitable detector for this analysis; however, ORCA could also be employed, particularly for the energy-independent \ac{qd} model. 
In Figure~\ref{fig:orca}, we show the expected deviations of neutrino fluxes from standard predictions under the constant decoherence model ($\Gamma = \gamma_0$), as the models with positive energy dependence are not relevant in the low-energy regime, and assuming a fermion number of $n_\text{f} = 4$. The plot suggests that we anticipate the signal to be within the energy range below 20 GeV, with a particular focus on below 10 GeV. This is the energy region for which ORCA is specifically optimised.
Additionally, an analysis using DeepCore \cite{DeepCore} would also be feasible, as it has an energy threshold of about 10 GeV.
Plots for ARCA are not shown as the expected signal region is similar to the IceCube one (see Figure~\ref{fig:signedchi2_true}).

\section{Summary and Discussion}
\label{sec:dis}
This work examines the impact of \acf{qg} on atmospheric neutrino oscillations, considering the presence of additional hypothetical dark fermions.
Since we do not assume an underlying theory of $\ac{qg}$, we adopt an effective description in the framework of open quantum systems.
This is based on the assumption that neutrino flavor eigenstates become entangled with unknown \ac{qg} \acf{dof} during propagation.
Since these \ac{dof} can not be directly probed with current experimental methods, only the flavor subsystem is observable. Consequently, decoherence arises in the neutrino flavor subsystem due to the entanglement between the observed and unobserved \ac{dof}.

A central assumption in this analysis is that \ac{qg} maximally violates the conservation of quantum numbers associated with global symmetries
and that it only conserves unbroken gauge quantum numbers as well as angular momentum and energy.
Due to the aforementioned decoherence effect an initially pure neutrino beam could then develop a non-zero component of potentially existing dark fermions.
Since \ac{qg} is assumed to maximally violate conservation of lepton flavor in the asymptotic limit all transition probabilities must approach $1 / n_f$,
where $n_f \geq 3$ is the total number of fermions. Possible candidates for dark fermions are WIMPs~\cite{Roszkowski:2017nbc}, sterile neutrinos~\cite{Dasgupta:2021ies}, and FIMPS~\cite{Westhoff:2023xho}. 

We consider several decoherence models differing only in the dependence of the decoherence parameter, $\Gamma \propto \gamma_0 E_\nu^n$,
on the neutrino energy $E_\nu$ with power-law $n$ and the total number of dark fermions $n_f$ determining the dimension of flavor space.
A statistical analysis, based on
%, based only on statistics and excluding systematic effects, i.e. 
publicly available \acf{mc} simulations, has been performed to determine the expected potential of the IceCube Neutrino Observatory to probe the \ac{qg} parameter $\gamma_0$.
The expected sensitivities have been determined by reweighting the public MC sample provided by IceCube taking into account the new decoherence effects as well as the systematic effects caused by the detector response and the initial neutrino flux model. The analysis also determines the specific energy ranges and zenith angles where sensitivities could be anticipated for IceCube.

The analysis was conducted for various total numbers of dark fermions.
The results confirm the expectation that the sensitivity increases with a higher number of fermions.
This demonstrates that, under the model assumptions, atmospheric neutrino experiments can in principle be used to discriminate between different new physics scenarios involving different numbers of dark fermions.
Moreover, combining astrophysical with atmospheric neutrino data can lead to further improvements potentially allowing for a simultaneous fit of all model parameters, including the number of fermion generations $n_f$, as astrophysical neutrinos are even more sensitive to the \ac{qg} effect due to their orders of magnitude larger travel distances.

% Beyond atmospheric neutrinos, astrophysical neutrinos are the most promising candidates for this study.
However, the currently limited statistics do not allow for a comprehensive analysis.
Meanwhile, combining atmospheric neutrino data from IceCube and KM3NeT detectors would be promising. Integrating data from both experiments would enhance the statistical power and address different systematic uncertainties, ultimately leading to a stronger potential for probing the QG parameters.

\section*{Acknowledgements}
A. Domi acknowledges the support from the European Union's Horizon 2021 research and innovation programme under the Marie Skłodowska-Curie grant agreement No. $101068013$ (QGRANT).

% ##################################################################################
% ################################## Bibliography ##################################
% ##################################################################################
\clearpage

\bibliographystyle{JHEP}
\bibliography{sn-bibliography}

\end{document}